\documentclass[12pt]{article}
\usepackage{slashed}

\usepackage{amsmath,amssymb}
\usepackage{bm}
\usepackage{epsfig}

\newcommand{\bpsi}{\bar{\psi}}

\begin{document}

\title{
 Quark and pion effective couplings from polarization effects
}

\author{ F\'abio L. Braghin
\\
{\normalsize Instituto de F\'\i sica, Federal University of Goias,}
\\
{\normalsize Av. Esperan\c ca, s/n,
 74690-900, Goi\^ania, GO, Brazil}
}

\maketitle
\begin{abstract}
  A  flavor  SU(2)   effective model  for pions and  quarks  is derived by
considering polarization effects departing from the  usual 
quark-quark effective  interaction induced by dressed  gluon exchange,
i.e.  a global color model for QCD.
For that,  the
quark field is decomposed into a  component  that
yields
 light  mesons and the quark-antiquark condensate,
being  integrated out
 by means of 
the auxiliary field method, 
and  another 
 component which yields  constituent quarks,
 which is  basically 
 a background quark field.
Within  a  longwavelength and  weak quark field 
 expansion
(or large quark effective mass expansion)
 of  a  quark determinant,
the   leading terms 
  are found up to the second order in
 a zero order derivative expansion,
 by  neglecting  vector mesons that are
considerably heavier than the pion.
Pions are considered in the structureless limit
and, besides the  chiral invariant terms that 
 reproduce previously derived expressions, symmetry breaking terms
are also presented.
The
  leading  chiral quark-quark effective couplings are also found
corresponding to a NJL  and a vector-NJL couplings.
All the resulting effective coupling constants and parameters
 are expressed in terms of 
the current and constituent quark masses and of the coupling $g$.
\end{abstract}


\section{Introduction}

Hadron and nuclear structure and dynamics are 
ultimately ruled by Quantum Chromodynamics (QCD)
which, due to its intrincated structure, is not exactly solved 
with currently known analytical  methods
\cite{brambilla-etal}.
There is a large amount of works
   dedicated to establish connections  between 
 (low and intermediary energies) QCD and
 observable hadrons 
 which rely on the derivation and/or elaboration of effective models, either
formally or pheonomenologically
\cite{kleinert,ERV,E-R1986,Weinberg2010,PRC,williams-vujinovic,weinberg79,cpt1,cpt2,brown-zahed,simonov-prd2002,osipov-etal,colomb,gezerlis-etal,EFT,gA-latt,qcd-hadron,mckay-munczek,largeNc,thooft-Nc,manohar-georgi,derafael,lavelle-mcm,thomas,PRD-2013,PRD-2014,NJL}.
Hadron effective field  models and theories   must then 
be compatible with  the more fundamental QCD symmetries and structure.
These effective  models and theories  are expected to 
describe observations  already in the tree level or with first order  corrections
and 
the corresponding  effective parameters and coupling constants
might  be expected to be calculable from QCD grounds.
Several   effective models  for 
low energy  hadron
 structure and  interactions 
are formulated in terms of pions and 
 constituent quarks and  gluons, which 
 have shown to be a powerful  way to describe many 
hadron properties
\cite{lavelle-mcm,thomas}.
Currently there is a large effort by many groups to describe
nuclear properties by considering hadron effective models or theories,
few examples are given  in  
\cite{Weinberg2010,brown-zahed,EFT,manohar-georgi,derafael,lavelle-mcm,thomas,NJL,epelbaum-etal,thomas-etal},
 whose QCD content is more transparent.
Dynamical
chiral symmetry breaking 
(D$\chi$SB) is  an extremely important  QCD effect  to be 
taken into account
since  it yields  broad and well known 
  consequences in the hadron level 
 and a direct relation to
confinement is expected, 
for example in
\cite{brambilla-etal,alkofer-etal,greensite}.
The relevance of the corresponding 
 chiral condensate for hadron structure and dynamics is  widely recognized 
although  recently its location became a  controversial subject 
 \cite{in-hadron-1,reinhardt-weigel}.
When establishing relations between QCD and hadron phenomenology
it is of utmost importance to  understand each piece of the 
mechanisms and effects that provides the measurable quantities.
The fundamental effects that 
yield  hadron effective interactions/couplings  might help to select the 
most realistic and relevant couplings in a hadron model.
One of the most investigated sources of 
quark effective couplings 
is  instanton mediation \cite{thooft-inst,shuryak,schaffer-shuryak,simonov-prd2002}
and  polarization effects  have also been shown to 
 produce quark effective interactions  \cite{PRD-2013,PRD-2014}.

D$\chi$SB is to  be 
investigated and understood
in the QCD quark sector
andm although this program still needs theoretical developments and improvements,
one might address  particular known  limits of the full theory.
The following low energy
 quark effective global color model
\cite{PRC,ERV,E-R1986}
  will be considered:
\begin{eqnarray} \label{Seff}  
Z &=& N \int {\cal D}[\bpsi, \psi]
\exp \;  i \int_x  \left[
\bar{\psi} \left( i \slashed{\partial} 
- m \right) \psi 
\right.
\nonumber
\\
&-&  \left.
 \frac{g^2}{2}\int_y j_{\mu}^b (x) 
({\tilde{R}}^{\mu \nu}_{bc} ) (x-y) j_{\nu}^{c} (y) 
+ \bpsi J + J^* \psi \right] 
\nonumber
\end{eqnarray}
Where the color  quark current is 
$j^{\mu}_a = \bar{\psi} \lambda_a \gamma^{\mu} \psi$, 
 the sums in color, flavor and Dirac indices are implicit, $\int_x$ 
stands for 
$\int d^4 x$,
the kernel ${\tilde{R}}^{\mu \nu}_{bc}$ is a dressed gluon propagator.
Even if other terms   might arise from the 
non abelian structure of the gluon sector,
 the quark-quark induced
interaction 
   (\ref{Seff})  
should  be  part of the  quark effective action  for QCD. 
Although  simple one gluon exchange is known to 
not produce enough strength of quark interactions
such as to yield   D$\chi$SB
there are different effects  that  improve  such picture
and make the strength strong enough.
 For example
by considering gluons self interactions in the polarization tensor,
by modeling confinement
 or   corrections in 
the corresponding Schwinger Dyson approach already in the rainbow ladder
approximation,
 few examples are given  in Refs. 
\cite{williams-vujinovic,cornwall-2011,DS-1,kondo}.

The aim of this work is to derive a  flavor  SU(2)  low energy
 effective model
  when
 internal structure of pions
is not important by departing from the generating functional
 for the model (\ref{Seff}).
A quark field splitting 
into two components  is performed: 
one component whose composite excitations correspond
to light quark-antiquark mesons and the scalar condensate, and the
other that  remains as (constituent) background field quarks.
If these components  may 
correspond or not to low or high energy quark components
will not be discussed since this  procedure corresponds to 
 the one-loop background field method
\cite{background,SWbook}.
Vector mesons degrees of freedom will be neglected since 
they are considerably heavier than the pion and are not expected to 
contribute in the low energy limit.
The analysis 
of   problems related to the light vector mesons  is left for another work.
The quark and pion 
effective couplings  emerge from
an expansion of a quark determinant
which therefore  generates a series of interactions with progressively higher powers 
of quark bilinears
and of the pion field and their  derivatives.
Mesons will be considered in the punctual approximation
within  the zero order derivative expansion 
and this  basically 
reproduces the corresponding terms found previously in 
Refs. \cite{PRC,E-R1986} in the chiral limit.
In addition to that, chiral symmetry breaking terms due to the current quark mass
are also found  in the punctual pion limit.
Besides that,  leading  quark-quark effective couplings, such as the 
NJL coupling, are also exhibited,
being higher order in $1/N_c$ in agreement with Ref.  \cite{wittenNc}.
We also believe this approach might provide  insights for the investigation
of the longstanding problem of the convergence of the QCD 
effective action.
An   effective field theory (EFT)
 for low and intermediary  energies QCD
might be though as  enough to provide 
a reliable understanding of the corresponding processes with 
enough predictive power  to describe
 Nature at this level.
However, 
this picture becomes  more complete
if the more fundamental mechanisms
 that generate
all the terms of the corresponding EFT,  with their  effective 
 parameters,  are  found  or derived.
With the present work we hope to provide further insight into this program
by considering the global color model (\ref{Seff}).

The paper  is organized in the following way.
In the next section, with a  Fierz transformed version of  the above 
non local current-current quark effective interaction,
 the quark field is separated into two components.
One of these components  will be integrated out by considering
a set of auxiliary fields which generate quark-antiquark
meson fields and the chiral condensate.
The other component   is dressed by 
polarization effects.
Auxiliary fields that correspond to  vector mesons
will be  neglected since they are considerably 
heavier than pions and do not contribute in the low energy regime.
With a chiral rotation in Sect. \ref{sec:chiral-rot},
 the non linear realization of chiral symmetry is 
introduced  in terms of   covariant derivatives.
The 
 expansion of the quark  determinant is exhibited  in Section \ref{sec:expansion},
where the effective  coefficients of the  leading terms of quark couplings   are presented
up to the second order and  the pion sector up to the fourth order.
In the last Section there is a summary  and discussion.

\section{ Flavor structure and auxiliary fields }
\label{sec:two-Q}

 The departure point is therefore expression (\ref{Seff}),
and the kernel $\tilde{R}^{\mu \nu}_{ab}$
 can be written  in terms of 
transversal and longitudinal components such that:
 \begin{eqnarray}
\tilde{R}^{\mu\nu}_{ab} = \delta_{ab} \left[
 R_T \left( g^{\mu\nu} - \frac{\partial^\mu \partial^\nu}{\partial^2}
\right) 
+ R_L \frac{\partial^\mu \partial^\nu}{\partial^2} \right].
\end{eqnarray}

With a Fierz transformation \cite{PRC,ERV,NJL} by 
selecting only  the color singlet terms,
the quark interaction above  can be written
in terms of 
 bilocal quark bilinears,
$j_i^q(x,y) =  \bpsi (x) \Gamma^q \psi (y)$ where 
$q=s,p,v,a$
and  
$\Gamma_{s} = I_{2} . I_4$ (for the 2x2 flavor and 4x4  identities),
 $\Gamma_{p} = i  \gamma_5 \sigma_i$,
$\Gamma_{v}^\mu =  \gamma^\mu \sigma_i $ and
$\Gamma_{a}^\mu =    i \gamma_5 \gamma^\mu  \sigma_i $,
where   $\sigma_i$ are the flavor SU(2) Pauli matrices.
The resulting non local  interactions are the following:
\begin{eqnarray} \label{fierz4}
 \Omega &\equiv& 
g^2 j_{\mu}^a (x) \tilde{R}^{\mu\nu}_{ab} (x-y) j_{\nu}^b (y)
\nonumber
\\
&\to&
 \alpha g^2 
\left\{ 
\left[ j_S (x,y)  j_S(y,x) + j_P^i(x,y)  j_P^i(y,x)  \right] R  (x-y)
\right.
\nonumber
\\
&-& \left.
    \frac{1}{2} \left[ j_{\mu}^i (x,y) j_{\nu}^i (y,x) 
+  {j_{\mu}^i}_A (x,y)  {j_{\nu}^i}_A (y,x)
\right]  \bar{R}^{\mu\nu} (x-y) \right\},
\end{eqnarray}
where 
$i,j,k=0,...(N_f^2-1)$ and 
$\alpha = 
 8/9$ for  SU(2)  flavor.
The kernels above
 can  be written as:
\begin{eqnarray} \label{Rbar-Rbar}
 R (x-y)  &\equiv&  R = 
3 R_T  + R_L,
\nonumber
\\
 \bar{R}^{\mu\nu} (x-y)  
&=&  g^{\mu\nu} (R_T+R_L) + 
2 \frac{\partial^{\mu} \partial^{\nu}}{\partial^2} (R_T - R_L).
\end{eqnarray} 
The local limit  
 yields the Nambu Jona Lasinio (NJL) and vector NJL couplings.
The coupling constants are roughly, for massless gluons,
 $G \sim \frac{g^2}{\Lambda_{QCD}^2}$  \cite{A-E} 
 or
 $G \sim \frac{g^2}{M_G^2}$ for non zero effective gluon mass 
being
comparable in any case
\cite{PRD-2013,NJL,njl-gluon-propag,qcd-njl,qcd-njl-2,1002.4151}.
These couplings are 
 of the order of
 $G \sim g^2  \sim 
1/N_c$.

The quark field   
will   be splitted
such as to preserve chiral symmetry
 into a  
component,  $(\psi)_2$, that yields the scalar condensate
and whose (composite) excitations correspond to 
(quark-antiquark)
 light mesons,
and another component, $(\psi)_1$, that will be associated to 
 constituent quark.
If the usual shift were performed with a background fermion field,
 for $\psi \to \psi + \psi_1 $
 and 
$\bar\psi  \to  \bar\psi + \bar{\psi}_1$,
 additional contributions of higher order would emerge.
This  field separation by means of the bilinears 
yields rather
 the background field method 
in the one-loop level \cite{background,SWbook}.
The  shift in the bilinears also  produces automatically chiral invariant structures, 
being however that quarks that are integrated out
 also provide    bound light meson states and the quark-antiquark condensate.
Physically, this shift of bilinears  can also be  
 associated  to the 
quark-antiquark  states built with auxiliary fields.
Quark  bilinears  will therefore be written as:
\begin{eqnarray} \label{split-Q} 
\bpsi \Gamma^q \psi \to  t_2  (\bpsi \Gamma^q \psi)_2 + t_1  (\bpsi \Gamma^q \psi)_1,
\end{eqnarray}
where $t_1$ and $t_2$  are  constants that can be set to one at the end and 
that help to understand the role of each of these components
in the resulting model.
According to this separation, 
the     current-current quark  interactions can be written in three parts:
$\Omega \to  t_1^2 \Omega_1 +  t_2^2 \Omega_2 + t_1 t_2\Omega_{12}$
where $\Omega_{12}$  mixes   both components and $\Omega_1, \Omega_2$ stand 
for the terms exclusive  to each of the components.
$\Omega_{12}$  can be written as:
\begin{eqnarray}
\Omega_{12}  &\equiv&
\alpha  g^2 \left[
{j^S}_1 (x,y) R {j^S}_2  (x,y) + {j_i^P}_1(x,y) R {j_i^P}_2 (x,y)
\right.
\nonumber
\\
&& \left. 
+ 
{j^S}_2 (x,y) R {j^S}_1  (x,y) 
+ {j_i^P}_2 (x,y) R {j_i^P}_1 (x,y)
 \right]
\nonumber
\\
&-& 
\frac{\alpha g^2}{2} \bar{R}_{\mu\nu} \left[  {j_i^\mu}_1 (x,y)  {j_i^\nu}_2 (x,y)
+   {{j_i^\mu}_A}_1 (x,y)  {{j_i^\nu}_A}_2 (x,y)
\right.
\nonumber
\\
&&
\left.
+ {j_i^\mu}_2 (x,y)   {j_i^\nu}_1 (x,y)
+   {{j_i^\mu}_A}_2 (x,y)   {{j_i^\nu}_A}_1 (x,y)
\right].
\nonumber
\end{eqnarray}
The resulting ambiguity in this splitting will not be solved  here.
However it will be shown below  that it yields the 
ambiguity of determining the relative contribution of constituent quarks
and pions (or pion cloud) to describe hadron observables, in particular 
baryons \cite{Weinberg2010,kalbermann}.

\subsection{Auxiliary fields and pions}

To integrate out the component $(\bpsi \psi)_2$,
a  set of bilocal auxiliary fields (a.f.)
with the quantum numbers of the bilinears defined above 
  is introduced to linearize  
$\Omega_2$.
The generating functional is multiplied by a
collection of  normalized unity Gaussian
integrals of the a.f.,
and these fields are shifted by fermion  bilineares, 
preserving an unit Jacobian, such that 
all the terms  in $\Omega_2$ are canceled out.
These integrals, with the corresponding shifts, are given by:
\begin{eqnarray}
 1 &=& N \int D[S] D[P_i]
 e^{- \frac{i}{2 } t_2^2
\int_{x,y}   R  \alpha  \left[ (S - g   j^S_{(2)})^2 +
(P_i -  g    j^{P}_{i,(2)} )^2 \right]}
\nonumber
\\
&&
 \int 
D[V_\mu^i]
 e^{- \frac{i}{4 } t_2^2
\int_{x,y} {\bar{R}^{\mu\nu}} \alpha
\left[ (V^i_{\mu} -  g    j_\mu^{i,(2)}) (V^i_{\nu} -
 g   j_\nu ^{i,(2)} )
\right]
}
\nonumber
\\
&&
\int  D[\bar{A}_\mu^i] 
 e^{- \frac{i}{4 } t_2^2
\int_{x,y} {\bar{R}^{\mu\nu}} \alpha
\left[ 
  ( \bar{A}^i_{\mu} -  g    {j_\mu^{i,(2)}}^A ) ( \bar{A}^i_{\nu} -
 g   {j_\nu^{i,(2)}}^A )
\right]
}.
\end{eqnarray}
The bilocal a.f. are $
S(x,y), P_i(x,y), V_\mu^i(x,y)$ and $\bar{A}_\mu^i(x,y)$ 
and  the corresponding 
shifts  (with unity Jacobian) were  given by the following  bilocal bilinears: 
\begin{eqnarray} \label{currents}
& j_S = \bpsi (x) \psi (y),
\;\;\;\;\;
& j_P^i = \bpsi (x) i \gamma_5 \sigma_i \psi (y),
\nonumber
\\
& j_\mu^i = \bpsi (x) \sigma_i \gamma_\mu \psi (y),
\;\;\;\;\;
& j_\mu^{i,A} = \bpsi (x) \sigma_i  i \gamma_5 \gamma_\mu  \psi (y).
\end{eqnarray} 
 The resulting  effective Lagrangian  is given by:
\begin{eqnarray}
{\cal L}_1 &=& 
\int_y  t_2  \bar{\psi}_2 \left[ ( i \gamma \cdot \partial
- m) \delta_{x,y} 
+ t_2 \alpha  g  R  (S   + i \gamma_5  \sigma_i P_i   )
+ \frac{t_2}{2 }
 \alpha g {\bar{R}^{\mu\nu}}
( V^i_{\mu} \gamma_\nu
+   \bar{A}^i_{\mu}  i \gamma_5 \gamma_\nu
 )
\right.
\nonumber
\\
&& \left.
+    t_1 \alpha g^2 \sum_q R^q  (x,y) \Gamma_q   j^q (x,y)
\right] \psi_2
+ t_1 \bar{\psi}_1  \left(  i \gamma \cdot \partial
- m \right) \psi_1
\nonumber
\\
&&
 - \frac{g^2 t_1^2}{2}
\int_y j_{\mu}^{b,(1)} (x)  \tilde{R}^{\mu \nu}_{bc} (x,y) j_{\nu}^{c,(1)} (y) 
- \frac{t_2^2 \alpha}{2 }
\int_{y}  \left\{ R    \left[ S^2 +
P_i^2 \right]
+ \frac{1}{2 }
 {\bar{R}^{\mu\nu}} 
\left[ V^i_{\mu} V^i_{\nu}  +
\bar{A}^i_{\mu}  \bar{A}^i_{\nu} 
\right] \right\},
\end{eqnarray}
where  terms of the form $ R^q (x,y) \Gamma_q  j^q (x,y)$
 correspond to the terms from $\Omega_{12}$,
being $R^q =  ( R, \bar{R}^{\mu\nu} )$ 
 with the corresponding 
operators $\Gamma^q$ from the bilinears.
In this expression the inverse Fierz transformation
was performed for the terms in $t_1^2 \Omega_1$
which was written as  a current-current effective interaction again.
By integrating out the quark field $\psi_2$,
in the limit of zero quark field $\psi_1$ (or $t_1=0$)
and zero quark mass,
the resulting model is the same as the  model presented
and investigated in Refs. \cite{kleinert,E-R1986,PRC} in the flavor 
SU(3) version.
Therefore the resulting pion sector has the same structure.
It  has also  been shown that a chiral rotation
in the measure of the generating functional
yields a  Wess Zumino term
\cite{E-R1986,PRC}.
Since the pion sector obtained in these works
is the same
as the one obtained in the present article the calculation will not
be exhibited here.

From this  non local theory ,
the local meson fields  are defined by means of 
a  formal 
expansion of the bilocal a.f. $\phi_q(x,y)$
on a local meson field basis  $M_{k,q}$
which  is given by: 
\begin{eqnarray} \label{nonloc-local}
\phi_q (x,y) = \tilde{F}_q (x-y) + \sum_k M_{k,q} \left(\frac{x+y}{2}
\right) F_{k,q} (x-y) ,
\end{eqnarray}
where $F_{k,q}$  are the form factors
 associated to the
corresponding $k=0,1,2...$-meson excitation of the channel  $q$.
 $\tilde{F}_q$ (for $q=s,p,v,a$) correspond to the 
translational invariant  vacuum functions  for each of the 
channels $q$ and  $M_{k,q}$  are the  local meson fields,
being that $k > 0$ correspond to all meson excitations
in the corresponding quark-antiquark channel $q$. 
The aim of the present work is to obtain  a local meson-quark 
effective model for the low energy regime in which internal structure of mesons 
is not relevant. 
Therefore only the local limit of 
these form factors  will be considered, and it will be written as:
$F_{k,q} (x-y) R^{q}(y-x) \simeq F_{k,q} \delta (x-y)$.
This limit yields punctual mesons.
Besides that, only  the  lowest quark-antiquark
scalar and pseudoscalar states (mesons) should contribute,
 i.e. $k=0$ for  the local meson fields denoted by 
$M_{k=0,q}=s,p_i,v_\mu^i,a_\mu^i$ since
higher quark-antiquark  excited states are heavier and only contribute for 
(relatively) higher 
energy
processes.
Finally,
the   vector/axial (quark-antiquark) 
mesons do not contribute for low energy regime
 since
their masses are considerably higher than the pion mass.
Even if  the auxiliary fields for  vector mesons  were considered, 
their structure and dynamics 
could receive contributions from constituent quarks and pion cloud,
inducing an ambiguity in their description similar to the one that 
will be found for baryons.
This problem however is outside the scope of this work.
Moreover,
 vector mesons are known to give rise to corrections for the 
Skyrme terms (fourth order pion couplings $c_1, c_2$ found below)
 \cite{vector-meson-skyrme}
 and therefore to some extent
their contribution 
 can be incorporated   by
redefining the fourth order pion terms.
  Chiral transformations mix scalar and pseudoscalar fields
and therefore,
 in the limit of small current quark masses,
one must have $F_{0,s} = F_{0,ps} =  F$. 
From here on, only the  local punctual  
meson fields leading terms will be considered.
This will produce  the correct punctual meson limit of the  previously derived
 low energy pion effective  couplings 
 \cite{PRC,ERV,E-R1986,simonov-prd2002}.

Consider the following  terms from the
 quark and auxiliary fields
 interaction, $\bpsi_2(x) \Xi (x,y) \psi_2 (y)$ where:
\begin{eqnarray} \label{q-meson-term}
\Xi (x,y) =
g \alpha 
\left\{ F_{0,0} (x-y)  R  \left[ S (z)  + 
P_i  (z) i \gamma_5 \sigma_i   \right] 
\right.
\nonumber
\\
-\left. 
 \frac{\gamma_\nu \sigma_i}{2}
\bar{R}^{\mu\nu}  \left[  F_{0}^{v} (x-y)  V_{\mu}^i (z)
+  i  \gamma_5 
 F_{0}^{a} (x-y) \bar{A}_{\mu}^i (z) \right]
  \right\}
\end{eqnarray}
where 
$z=\frac{x+y}{2}$ that reduces to  $z=x$ due to the structureless mesons
approximation,
and, in the absence of the heavier vector mesons, it 
reduces to:
\begin{eqnarray} \label{scalars-q}
\Xi (x,y) &\to& \Phi_L (x,y) \simeq  F 
  \left(s  + p_i \gamma_5 \sigma_i  
\right) \delta (x-y) 
\nonumber
\\
&\equiv&  \tilde{\Phi}_L \delta (x-y).
\end{eqnarray}
Therefore, the quark $\psi_2$
 and  meson  coupling 
in terms of  $\Xi \sim  \tilde\Phi  \delta^4 (x-y)$  can be written as:
$ 
 t_2^2  \; \bpsi_2 \Phi_L (x-y) \psi_2$,
which corresponds to the linear realization of chiral symmetry.
The canonically  normalized definition 
of the pion field becomes:
$\vec{\xi} = F \vec{p}$.

\subsection{ Chiral rotation}
\label{sec:chiral-rot}

The non linear representation for chiral symmetry
can be obtained as described below.
The scalar field is  frozen
and then, by performing  a chiral rotation,  only the pion field and its  (covariant)
 derivative remain \cite{chiral-rot,weinberg1968,SWbook}.
This can be done by constraining the scalar and pseudoscalar fields  to 
the chiral radius,
$1 = s^2 + \vec{p}^2$ which yields $s = \sqrt{ 1 - \vec{p}^2}$.
Now we note there is a freedom to define the pion and quark fields and derivatives
related among each other by chiral rotations.
The quark free  terms and its   coupling   to (scalar and pseudoscalar) mesons
are  given by:
\begin{eqnarray} \label{sigmamodel}
t_2 \; \bpsi_2 \left[  i \gamma \cdot  \partial
 -  m  +
 t_2 \Phi_L \right] \psi_2
\nonumber
\\
  = t_2 \; \bpsi_2 \left[ i \gamma \cdot  \partial
 - m +
 t_2  F ( s + i \gamma_5 \vec\sigma \cdot \vec{p} ) \right]
\psi_2,
\end{eqnarray}
Quark and scalar and pseudoscalar fields can be  redefined  as 
\cite{chiral-rot,weinberg1968,SWbook}:
\begin{eqnarray} \label{quark-D}
s =  \frac{1 - \vec{\pi}^2}{1+ \vec{\pi}^2} ,
\hspace{.3cm}
 {p}_i = \frac{2 \pi_i}{1 + \vec{\pi}^2} ,
\hspace{.3cm}
\psi &=& \frac{(1  - i \gamma_5 \vec{\sigma}
 \cdot \vec{\pi})}{\sqrt{1+ \vec{\pi}^2}}  \psi'.
\end{eqnarray}
In the resulting  non linear realization of chiral symmetry
the above Lagrangian  terms (\ref{sigmamodel}) can be written as:
\begin{eqnarray}
 \bpsi_2' 
\left[
 i \gamma \cdot  \partial
-   m^*
+
   \gamma^\mu \vec\sigma \cdot \left(
\frac{\partial_\mu \vec\pi}{1+ \vec\pi^2} i \gamma_5 +
i 
\frac{\vec{\pi} \times \partial_\mu \vec{\pi}}{1+{\vec{\pi}}^2} 
\right)
\right.
\nonumber
\\
\left. 
+ 4  m  
\left( 
\frac{\vec\pi^2}{1+ \vec{\pi}^2}
-  \frac{\epsilon_{ijk} \sigma_k \pi_i \pi_j }{1+ \vec{\pi}^2}
\right)
\right]
\psi_2',
\end{eqnarray}
where
it was used that 
$ \sigma_i \sigma_j = \delta_{ij} + i \epsilon_{ijk} \sigma_k
$,
and  it has been set
$$t_2=1.
$$
The last two  terms in this expression correspond to the chiral symmetry breaking term
from the current quark mass.
These terms
 yield the terms proportional to the pion mass or to powers of $\vec\pi^2$ in the resulting
effective model.
To improve the notation two covariant derivatives are defined  as:
\begin{eqnarray} \label{chiral-deriv}
{\cal D}_\mu  \vec{\pi} \equiv \frac{\partial_\mu {\vec\pi}}{(1 + \vec{\pi}^2)},
\nonumber
\\
\bpsi_2  \partial_\mu \psi_2 \to
\bpsi_2'   D_{\mu}^{c} \psi_2' \equiv 
\bpsi_2'  \left( \partial_\mu   +  i  \vec{\sigma} \cdot 
\frac{\vec{\pi} \times \partial_\mu \vec{\pi}}{1+{\vec{\pi}}^2} 
\right) \psi_2'.
\end{eqnarray}
The canonically  normalized pion field corresponds to 
$\vec{\pi}'= \vec{\pi} F$.
From here on, the new definitions of pion and quark field will be used by writing simply
$\pi_i$ and $\psi_2$ respectively.
This redefinition of the fields however induces 
 a non trivial   change in the 
functional measure with terms that do not depend on this pion
covariant derivative. 
These terms are of higher order in the pion and quark fields,
  therefore they
 should be less important 
from a dynamical point of view.
This subject  will not be addressed further  in the present work, and therefore 
the  Jacobian will not be exhibited and discussed here.

A different parameterization of
the non linear realization   can be used  for the pseudoscalar 
fluctuations
around the vacuum to rewrite expression
(\ref{sigmamodel}), as discussed in Refs. \cite{E-R1986,PRC},
 by means of :
\begin{eqnarray} \label{Linear-NLinear}
\Phi_L \to \Phi_{NL}
&=&  
F \left( P_R U + P_L U^\dagger \right),
\end{eqnarray}
where  $U = e^{i \vec{\sigma} \cdot \vec{\pi}}$ 
and 
  $P_{R,L} = (1 \pm \gamma_5 )/2$
are the chirality projectors.
These expressions
allow to rewrite the pion sector
in the standard shape of
 Chiral Perturbation Theory.

\subsection{ Integrating out  quarks }
\label{sec:integrating}

By integrating out the 
 component $(\bpsi \psi)_2$
 the following non linear (non local) effective action 
for quarks $(\bpsi \psi)_1$ and pions  is obtained:
\begin{eqnarray} \label{Seff-q4}  
S_{eff}  &=& - i \; Tr \log \left\{
({S_0^c})^{-1}
+   
  \Phi_N 
- \alpha t_1 g^2 \bar{R}^{\mu\nu} \gamma_\mu  \sigma_i \left[
 (\bpsi \gamma_\nu  \sigma_i \psi)_1
+  i \gamma_5   (\bpsi 
i \gamma_5 \gamma_\nu  \sigma_i \psi)_1 \right]
\right.
\nonumber
\\
&+& 
\left.
 2   \alpha t_1 g^2 R
 \left[  (\bpsi \psi)_1 
+ i  \gamma_5 \sigma_i  (\bpsi i \gamma_5  \sigma_i \psi)_1 \right]
+  4  m \frac{(\vec\pi^2   -  \epsilon_{ijk} \sigma_k \pi_i \pi_j )}{1+ \vec{\pi}^2}
 \right\}
- 
\int 
 {\cal L}_{2}  ,
\end{eqnarray}
where 
the following relation was used: $\det (A) = e^{Tr \ln A}$
and where  $Tr$ stands for traces of discrete internal indices 
and integration of  spacetime or momentum 
  coordinates for the quark component
$\psi_2$.
The following kernel has been defined:
\begin{eqnarray}
({S_0^c})^{-1} \equiv
   (  i \gamma_{\mu} \cdot { D_{c}}^{\mu} 
-  m^* )    .
\end{eqnarray}
The contribution of the pion covariant derivative was 
written as: 
\begin{eqnarray}
\Phi_N = i \gamma_5 \gamma^\mu \vec\sigma \cdot {\cal D}_\mu \vec{\pi}
\end{eqnarray}
and the remaining terms for the first component of the quark field given by:
\begin{eqnarray} \label{free-q}
{\cal L}_2 &=&
 t_1 \bar{\psi}_1  \left(  i \gamma \cdot \partial
- m \right) \psi_1
 - \frac{g^2 t_1^2}{2}
\int_y j_{\mu}^{b,(1)} (x)  \tilde{R}^{\mu \nu}_{bc} (x,y) j_{\nu}^{c,(1)} (y) ,
\end{eqnarray} 

The  a.f. vacuum expected values
can be found from their gap equations.
However these equations must be found from the 
 effective action  in terms of the fields $s,p_i$, i.e. 
by integrating out quarks $\psi_2$
without doing the chiral rotrations of the last section.
These saddle point equations correspond to:
\begin{eqnarray} \label{gaps}
\left. \frac{\partial S_{eff}}{\partial \phi_i}
 \right|_{ [\phi_j^{(0)}=s^{(0)},p_i^{(0)},.. ]}
&=&0.
\end{eqnarray}
This set of equations corresponds basically to the usual set of gap equations of
the NJL model being that 
 in the Nambu Goldstone mode only the scalar field has 
a non zero expected value in the vacuum.
This  provides the only contribution 
to the quark effective mass that constitutes pions
and that condense into the chiral condensate,
$m^* = m + g s^{(0)}$.
These gap equations were solved,
for example, 
 in a model with a  very simplified gluon propagator
 in Ref. \cite{PRD-2014}  in terms of an ultraviolet  Euclidean cutoff.
When a particular  gauge is choosen for $R$ and $\bar{R}$, the gauge 
fixing parameter can be determined by 
a condition of gauge independence such as:
$\frac{\partial S_{eff}}{ \partial \lambda } = 0$.
All
the quantities in the effective action found below for quarks and pions,
 and also the gap equations above,
 depend basically on 
the original QCD Lagrangian parameters: u-d current quark masses, gauge coupling   $g$,
 a gauge fixing parameter  $\lambda$.

 By 
factorizing $\log ((S_0^c)^{-1})$ in the determinant,
with the quantity $\chi =   \gamma^\mu \vec{\sigma} \cdot 
\frac{\vec{\pi} \times \partial_\mu \vec{\pi}}{1+{\vec{\pi}}^2}$,
this term  can be written as:
\begin{eqnarray} \label{factorized}
S_{d2}'  = -   i \; Tr \; \log \left[   i \gamma \cdot  { \partial}
- m^* - \chi   \right]   = C_0 + S_{d2} 
\nonumber
\\
=  
-   i \; Tr \; \log \left[   i \gamma \cdot  { \partial}
- m^*   \right]   -   i \; Tr \; \log \left[  1 -   S_0 \chi   \right]   ,
\end{eqnarray}
where $S_0 = S_0^c (\pi_i=0)$,
being that the first term
($C_0$)
 becomes a multiplicative constant factor in the generating functional
and the second one 
can be expanded for weak pion field or  in a longwavelength  expansion.
The first order term is zero, and the expansion  can be written as:
\begin{eqnarray} \label{Sd2}
S_{d2}  =  i \sum_n \frac{1}{n}Tr \left( - S_0 \chi \right)^n.
\end{eqnarray}
 The expansion of $S_{d2}$  yields terms of higher order in the pion field than 
the expansion of the pion sector of  the remaining part of the determinant.
The quark determinant can then be written as:
$$S_{det} = S_d + S_{d2},
$$
where the main part   can  now written as:
\begin{eqnarray} \label{largeN}
S_{d} &=& - i \; Tr \log \left[ 1+   
S_0^c  \left( 
\tilde\Phi_N  
+ 4  m \frac{(\vec\pi^2   -  \epsilon_{ijk} \sigma_k \pi_i \pi_j )}{1+ \vec{\pi}^2}
+
g^2 \alpha t_1  \sum_q R_q \Gamma_q \bpsi \Gamma_q \psi \right)
\right],
\end{eqnarray}
 The pion coupling in $D_\mu^c$ also produces further interactions
in the expansion as shown below, 
however the most 
relevant one is  the first order term.
This determinant will be expanded in the longwavelength limit
(low pion momenta) and for weak $\psi_1$  quark field (or for small $g^2$).
This expansion is also equivalent to 
a large quark mass ($m^*$) zero order derivative expansion  \cite{chan}.

\section{ Effective quark and pion couplings   }
 \label{sec:expansion}

By neglecting 
vector  meson  fields
the expanded determinant
can be written as:
\begin{eqnarray}   \label{exp-1}
S_{d} &\simeq& i \;
Tr \sum_n
c_n 
\left\{   
S_0 
\left[
 {\Phi_N}   
+
2 K_0  R(x-y)  \left[
 (\bpsi (x) \psi (y))  
+
 \gamma_5     \sigma_i  
 (\bpsi (x)  \gamma_5  \sigma_i \psi (y) ) \right]
\right. \right.
\nonumber
\\
&-&
\left. \left.
K_0  \bar{R}^{\mu\nu} (x-y) 
\gamma_\mu \sigma_i  \left[
\bpsi (x)  \gamma_\nu  \sigma_i \psi (y)
+  i \gamma_5       \bpsi  (x)
i \gamma_5 \gamma_\nu \sigma_i  \psi (y) \right] 
\right. \right.
\nonumber
\\
&+& \left. \left.
 4  m \frac{\vec\pi^2   -  \vec\sigma \cdot \vec\pi \times \vec\pi }{1+ \vec{\pi}^2}
\right]
\right\}^n
\end{eqnarray}
where: 
 $c_n = \frac{(-1)^{n+1} }{n}$, and $K_0 = \alpha g^2  t_1$.
All the terms of this expansion will be calculated in the zero order 
derivative expansion.
Besides that,  only the leading terms in the pion derivative
will be shown, i.e., terms of higher order in  $(\partial^n \vec{\pi})$  ($n\geq2$)
 will be neglected.

Many terms in this expansion 
are zero due to the traces of Dirac and  Pauli
matrices.
The only non zero first order terms 
yield   a correction to the  quark $(\bpsi \psi)_1$
  mass 
and a  pion mass term. 
They are respectively the following: 
\begin{eqnarray} \label{Meff}
{\cal L}_1 &=& 
t_1  \Delta  m^* \; 
(\bpsi  \psi)_1
-
 M_\pi^2 F^2
\frac{\vec\pi^2}{1+ \vec{\pi}^2}
,
\end{eqnarray}
%
and where
 these   masses
were defined as:
\begin{eqnarray} \label{Meff-2}
\Delta m^*
&=&  - i  \;  N_c    \alpha  g^2 \;  Tr'  \; {S}_{0}   R,
\\
\label{Mpi}
M_\pi^2 &=& 
 i N_c \frac{1}{F} \; Tr'  \; S_0  m ,
\end{eqnarray}
where $Tr'$, from here on,
 corresponds to  traces in all internal and spacetime (or momentum)
 indices except color.
With the help of the gap equation for the scalar field (\ref{gaps}),
it  can be noticed that  expression (\ref{Mpi}) 
corresponds to the
 Gell Mann Oakes Renner relation: $M_\pi^2 F^2 = - <\bar{q} q > m_q$.
In this expression  it is seen that 
pion mass  becomes zero in  the chiral limit 
($m_q=0$), i.e. the Goldstone theorem is satisfied.
i.e. in the limit of structureless pions. 
If pion structure had not been neglected, expression
above with the corresponding gap equation for the quark (and eventually gluon) 
propagator
 yield
a  rainbow ladder Schwinger Dyson  
and Bethe Salpeter equations which had been solved previously
\cite{SD-goldstone,BS} and that is outside the scope of this work.

\subsection{Manohar and Georgi expansion}

By neglecting 
any coupling to  gluons
the  series above yields
 terms of the following general  type \cite{manohar-georgi} for $C=0$:
\begin{eqnarray}
I_{ABCD} &\sim& 
{\cal C}_{A,B}^{C,D} 
\left( \frac{\pi}{f} \right)^A
\left( \frac{\bar{\psi} \Gamma \psi}{f^2 \Lambda} \right)^B
\left( \frac{{\cal G}[A_\mu^{1,a}]}{\Lambda} \right)^C
\left( \frac{p}{\Lambda} \right)^D
\end{eqnarray}
which was   considered by Manohar and Georgi  
where ${\cal C}_{A,B}^{C,D}$ are the  coefficients to be calculated,
$({\cal G} [A_\mu^{a,(1)}])^C$ are gauge invariant combinations of 
$A^{a}_\mu$  which are not considered in this work ($C=0$),
with the
chiral invariant combinations of interacting  mesons/pions or quarks ($A,B$)
with momenta of order $D$.
Basically the  $n-th$ term in the expansion above corresponds to
 $A+B = n$ .
Momentum dependence will be considered solely
for the pion field in this work.

\subsection{ Second order terms}

 There is a first order term of the expansion above
that yields   a second order 
pion-quark coupling if the kernel $S_0^c$   is also expanded 
for the pion coupling $\chi$ in the first order,
i.e.  similarly to the expansion (\ref{Sd2}).
The resulting  term
  is the same as the one 
 emerging from the chiral rotation for the component $\psi_2$
by considering the zero order derivative expansion.
It can be written as:
\begin{eqnarray}   \label{deriv-1}
{\cal L}_{2d\pi} &=&  i \;  t_1 g_{\pi d \pi}
\frac{\vec{\pi} \times \partial^\nu \vec{\pi}}{1+{\vec{\pi}}^2}  
\cdot (\bpsi   \gamma_\nu  \vec{\sigma} \psi) ,
\end{eqnarray}
where it was defined the following coupling
constant:
\begin{eqnarray}  \label{gpidpi}
g_{\pi d\pi} g_{\rho \nu} \delta_{ij} &=& 
 i   g^2   \alpha N_c \;  Tr' \; S_{0}^2 \sigma_i 
\gamma_\rho
 \bar{R}_{\mu\nu}  \gamma^\mu \sigma_j .
\end{eqnarray}
In expression (\ref{gpidpi}) the following trace properties  must be used:
\begin{eqnarray}
tr (\sigma_i \sigma_j) &=& 2 \delta_{ij},
\nonumber
\\
tr( \gamma_\mu \gamma_\nu ) &=& 4 g_{\mu\nu},
\nonumber
\\
tr( \gamma_\mu \gamma_\nu \gamma_\rho \gamma_\sigma)
&=&
4 ( g_{\mu\nu} g_{\rho\sigma} + g_{\mu \sigma} g_{\nu \rho}
- g_{\mu\rho} g_{\nu \sigma} ), 
\end{eqnarray}
and, besides that,   rotational invariance for
the traces in  spacetime or momentum coordinates.
These properties must be used 
in    the second order terms of the expansion, 
in particular those for the Lorentz indices and rotational invariance. 
If  $ g_{\pi d \pi}=1$
 this coupling is precisely that coupling 
from the chiral rotation in the quark covariant derivative 
(\ref{chiral-deriv}).

The other  second order terms 
 are those for 
 free meson terms, 
quark-meson and quark-quark effective interactions
and they  are presented below in the zero order derivative expansion.
The resulting  pseudoscalar kinetic term  is given by:
\begin{eqnarray} \label{freePI}
\Delta {\cal L}_{free mesons}=   
\frac{k_2}{2}
 {\cal D}_\mu \pi_i  {\cal D}^\mu \pi_i ,
\end{eqnarray}
where 
\begin{eqnarray}
k_2 g_{\mu\nu} \delta_{ij}
 =   i  \;   N_c     \;  Tr'  \; \sigma_i \sigma_j \;  \gamma_5
\gamma_\mu \gamma_5 \gamma_\nu {S}_{0}^2 .
\end{eqnarray}
This expression corresponds to the punctual pion limit of 
the complete expression investigated in Refs.  \cite{PRC,ERV,E-R1986},
being 
 $k_2 = f_{\pi}^2$.

The leading  quark-meson coupling  is the axial coupling
\cite{gA-latt,weinberg-gA,bira-dicus,broniowski}
that is  ${\cal O} (N_c^0)$.
This term is  given by:
\begin{eqnarray} 
\label{2nd-I}
{\cal L}_{2}^I  &=&  
- 2 g_A t_1  ({\cal D}_\mu {\pi}_i) 
   \bpsi   i \gamma_5 \gamma^\mu \sigma_i \psi
 ,
\end{eqnarray}
Where the axial coupling constant  
 was defined as:
\begin{eqnarray} \label{gA}
g_A  \delta_{ij} g_{\rho\mu}
 &=& 
 - \frac{ i}{2}   \alpha  g^2   N_c 
 \;    Tr'   \; \left[
  \gamma_\rho \gamma_\mu (-1) 
\gamma_5^2 \sigma_i \sigma_j 
\bar{R} {S}_{0}^2  \right]
,
\end{eqnarray}
where $\bar{R}=g_{\mu\nu} \bar{R}^{\mu\nu}$
and the factor $(-1)$ is due to the anticommutation of 
Dirac matrices.
This coupling constant  depends strongly
on the gluon propagator, and  all the effective 
quark couplings will have this   dependence.

Also, there is a symmetry breaking pion-quark term in the second order given by:
\begin{eqnarray} \label{sigma-term}
{\cal L}^{sb}_2 
&=&
t_1   c_{\sigma,sb}
 \frac{\vec{\pi}^2}{1 + \vec{\pi}^2} ( \bpsi \psi )
\end{eqnarray}
where the coefficient is given by:
\begin{eqnarray}
c_{\sigma,sb} =  i \; 4 g^2 \alpha  N_c \; Tr' \; m \; S_0^2 R.
\end{eqnarray}
Expression (\ref{sigma-term}) 
is   the sigma term interaction.

\subsection{Second order quark effective interactions}

The longwavelength  limit of the  leading chiral
quark-quark couplings contains coupling constants which  
are of the order of  $1/N_c$.
The corresponding  second order terms of the determinant expansion above
 are
 given by:
\begin{eqnarray} \label{4quark}
{\cal L}_{4q} &=& t_1^2 
\; 
  g_{NJL}^{h.o.}  \left[ ( \bpsi  \psi )^2
 +   (  \bpsi \sigma_i \gamma_5 \psi )^2  \right]
\nonumber
\\
&+&  t_1^2 
\;
 g_{vNJL}  
\left[ ( \bpsi  \sigma_i \gamma_\mu \psi )^2
 + ( \bpsi \sigma_i \gamma_\mu \gamma_5 \psi )^2
\right] .
\end{eqnarray}
For these expressions, the coupling constants were defined within the 
zero order derivative expansion  in the following way:
\begin{eqnarray}
g_{NJL}^{h.o.} &\to& 
-  2 i g^4 t_1^2  \alpha^2  \; Tr \; {{S}_{0}}^2  R^2
\nonumber
\\
&& \times [ ( \bpsi   \psi)^2  -   \sigma_i \sigma_j i^2 \gamma_5^2
 (\bpsi i \gamma_5 \sigma_j   \psi)
 (\bpsi i \gamma_5 \sigma_i   \psi)] 
\nonumber
\\
&=& 
t_1^2 \; 
 g_{NJL}^{h.o.} 
[ ( \bpsi   \psi)^2  +    (\bpsi \gamma_5 \sigma_j   \psi)^2 ]
\nonumber
\\
g_{vNJL} &\to&
- \frac{i}{2} g^4 t_1^2 \alpha^2  \; Tr \; 
 \bar{R}^{\mu\nu} \bar{R}^{\rho \sigma}  
  {{S}_{0}}^2    \sigma_i \sigma_j
\gamma_\mu \gamma_\rho
\nonumber
\\
&& \times [ (\bpsi \sigma_i \gamma_\nu \psi) (\bpsi \sigma_j \gamma_\sigma \psi)  - i^2
\gamma_5^2
(\bpsi i 
\gamma_5 \sigma_i \gamma_\nu  \psi )(\bpsi i \gamma_5 \sigma_j \gamma_\sigma  \psi )]
\nonumber
\\
&=& 
t_1^2  \;
g_{vNJL}
 [ (\bpsi \sigma_i \gamma_\nu \psi)^2  +
(\bpsi \gamma_5 \sigma_i \gamma_\nu  \psi )^2 ],
\end{eqnarray}
where:
\begin{eqnarray}  \label{NJL-couplings}
g_{NJL}^{h.o.} &=&
-  2  i g^4  \alpha^2  N_c
\;   Tr'  {S}_{0}^2   R^2 ,
\nonumber
\\
 g_{vNJL} \;  g_{\mu\rho} \delta_{ij}  &=&
- \frac{ i}{2} g^4   \alpha^2  N_c 
\;   Tr'
\; \gamma_\mu \gamma_\rho 
\sigma_i \sigma_j   \; {S}_{0}^2  R_{2v} ,
\end{eqnarray}
where 
$R_{2v} = 4(R_T+R_L)^2 + 8 R_T(R_T-R_L)$.
With the relative contribution of the longitudinal and transversal
components of the gluon propagator it is possible to obtain simple relations 
between these two effective quark coupling constants.
If 
it is  assumed  
$Tr'' (R_T^2  {S}_0^2) >>   
Tr'' ( R_T  R_L {S}_0^2)$ and $ >>  
Tr'' ( R_L^2 {S}_0^2)$   then:
\begin{eqnarray}
\frac{g_{NJL}^{h.o.}}{g_{vNJL}} =  4 \frac{ Tr''  {S}_0^2   [3 R_T +R_L ]^2 }{
Tr''( S_0^2  R_{2v})} \sim 3,
\end{eqnarray}
whereas for 
$Tr'' ( R_L^2 {S}_0^2 )  
 >> 
Tr'' ( R_T R_L  {S}_0^2) $
and 
\\
$>> Tr'' ( R_T^2 {S}_0^2)$
 it yields:
\begin{eqnarray}
\frac{g_{NJL}^{h.o.}}{g_{vNJL}} = 4 \frac{ Tr'' ({S}_0^2  [3 R_T +R_L]^2 )}{
Tr'' (S_0^2  [R_{2v} ] )} \sim  1.
\end{eqnarray}
These limits are  in  agreement with phenomenology
 \cite{sugano}. 
The  emerging quark-quark potential
is therefore  composed by several types of chiral invariant terms 
and this intrincated structure is expected 
from a confining theory \cite{alkofer-etal}.
The   expressions for  sixth and eighth order effective quark 
 interactions, found  in Ref.  \cite{PRD-2014},  are  reproduced
without an explicit form of the gluon propagator.
By rescaling these expressions according to 
the 't Hooft's  large $N_c$ scheme 
the $n$-quark effective interactions obtained from this
expansion   are seen to be  of the order of
$N_c^{1-n}$ in agreement with the QCD large $N_c$ expansion \cite{wittenNc}
although the present calculation includes a restricted class of diagrams for 
the model (\ref{Seff}).
 There are certainly  third and higher  order couplings, 
2-mesons-quark couplings 
and 
 meson couplings to two quarks which will not be presented here.
They are basically  approximatedly
 one order of $1/m^*$ (or $1/(m^*)^2$)  and also $1/N_c$ smaller
than the second order axial pion-quark  coupling shown above.

\subsection{ Pion sector}

The longwavelength expansion provides pion
self interactions in the lines of chiral perturbation theory.
This was discussed with details in Refs.\cite{PRC,ERV,E-R1986}
for flavor  SU(3).
There are several contributions in the fourth order pion 
derivative and higher order in the pion field
which also receives contributions from the derivative expansion.
Since  the main aim of the present work is to provide quark-pion couplings 
 only the zero order derivative expansion terms will be shown here.
This will not  provide all the leading terms of chiral perturbation theory
(up to the fourth order in pion momenta)
and the missing term is obtained from the  derivative expansion 
as shown below.
The chirally symmetric  third order pion self interactions are zero,
 in the zero order derivative expansion
  the fourth order pion self couplings from $S_d$
 are  given by:
\begin{eqnarray}   \label{exp-2}
{\cal L}_{4}^{meson} &=&
- \frac{ i}{4} 
Tr 
\left\{  
S_0
  \Phi_N
\right\}^4
\\
&=& - \frac{i}{4}
Tr 
\left\{ 
 \gamma \cdot  {\cal D}  \pi_i ( \gamma_5 \sigma_i) 
 \gamma \cdot  {\cal D}   \pi_k  ( \gamma_5 \sigma_k )
\gamma \cdot {\cal D}   \pi_l ( \gamma_5 \sigma_l  )
 \gamma \cdot    {\cal D}   \pi_m ( \gamma_5 \sigma_m )
{S}_{0}^4 \right\} .
\nonumber
\end{eqnarray}
The traces in Dirac and flavor  indices 
are given by:
\begin{eqnarray} \label{properties-4}
 tr ( \gamma^{\mu} \gamma^{\nu} \gamma^{\rho} \gamma^{\sigma} ) 
&=& 4
\left(  g^{\mu \nu} g^{\rho \sigma} + g^{\mu \sigma} g^{\nu \rho}  
- g^{\mu \rho} g^{\nu \sigma}
\right),
\nonumber
\\
tr ( \sigma^i \sigma^j \sigma^k \sigma^l ) &=& 
2 ( \delta^{ij}  \delta^{kl} + \delta^{il} \delta^{jk} - \delta^{ik} \delta^{jl}).
\end{eqnarray}
 These fourth order terms ${\cal O}(N_c)$
 can be written
as:
\begin{eqnarray} \label{I-4}
{\cal L}_{4}^{meson} =
-   
c_1 ({\cal D}_\mu \pi_i \cdot {\cal D}^\mu \pi_i)^2 
-  c_2  ({\cal D}_\mu \pi_i \cdot  {\cal D}_\nu \pi_i)^2 
,
\end{eqnarray}
 where  $c_1 = - 2 c_2$
in agreement with Ref. \cite{PRC}  in the limit of point-mesons for which:
$
c_1 = 2 i   N_c 
  Tr''   {{S}_0}^4$, where $Tr''$ stands for trace in spacetime or momentum   indices.
These are the most general chiral fourth order terms for single pion derivatives
\cite{weinberg79}.
Although it is outside the  scope of this work to present 
an extensive investigation of the resulting 
effective field theory, 
the leading symmetry breaking terms
from this expansion
 are presented 
below in the zero order derivative expansion.
 Up to the second 
 order in the current quark mass, 
the following   symmetry breaking contributions appear:
\begin{eqnarray} \label{2nd-sb} 
{\cal L}^{sb}_4
 &=&
V_2
\frac{\vec\pi^4}{(1+ \vec{\pi}^2)^2}
+ V_{3B} 
\frac{\vec\pi^2}{(1+ \vec{\pi}^2)}
\left( {\cal D}^\mu \vec{\pi} \cdot {\cal D}_\mu \vec{\pi} \right)
- V_{3C}
\frac{
\left( \vec{\pi} \cdot {\cal D}^\mu \vec{\pi}  \right)^2
}{(1+ \vec{\pi}^2)}
\nonumber
\\
&&  + V_{4B} \frac{(\vec\pi^2)^2}{(1+ \vec{\pi}^2)^2}
\left( {\cal D}^\mu \vec{\pi} \cdot {\cal D}_\mu \vec{\pi} \right)
+V_{4C} 
\frac{  (\vec{\pi} \cdot {\cal D}^\mu \vec{\pi})
(\vec{\pi} \cdot {\cal D}_\mu \vec{\pi})
}{(1+ \vec{\pi}^2)^2} 
,
\end{eqnarray}
where the following effective parameters 
(low energy coefficieints) 
 were defined:
\begin{eqnarray} \label{parameters-sb}
V_2 &=&  - i  N_c  16
\; 
Tr'' \; m^2 S_0^2
,
\\
V_{3B} g^{\mu\nu} \delta_{ij}
&=& 
  i   N_c \frac{16}{3}
\; Tr''
 \; m
\gamma_5 \gamma^\mu \gamma_5 \gamma^\nu
\sigma_i \sigma_j S_0^3
\nonumber
\\
&+& 
V_{3C} g^{\mu\nu} \delta_{ij}
,
\\
V_{3C} g^{\mu\nu} \delta_{ij} &=&
  i   N_c 4 
\; Tr''
 \; m
 \gamma^\mu \gamma^\nu
\sigma_i \sigma_j S_0^3
,
\\
V_{4B} g^{\mu\nu} \delta_{ij}&=&
 -  i N_c  4
\; Tr''
 \; m^2
\gamma_5 \gamma^\mu \gamma_5 \gamma^\nu
\sigma_i \sigma_j S_0^4
,
\\
V_{4C} g^{\mu\nu}
&=&
 -  i N_c  4
\; Tr''
 \; m^2
\gamma_5 \gamma^\mu \gamma_5 \gamma^\nu
 S_0^4
.
\end{eqnarray}
The next leading terms of the derivative expansion for the term
$V_2$ contributes for  other terms in $I^{sb}$.

\subsubsection{   
Rewriting the pion sector  
}

Let us rewrite the pion sector by 
considering the  parameterization given in  
expression (\ref{Linear-NLinear}).
For that, 
the derivative couplings must be extracted.
The   kernel  ${S}_{0}^N$
now  is a function of the  current  quark mass $M$.
The terms in the traces 
$Tr$ 
can be written in terms of a part diagonal in coordinate space
and another diagonal in momentum space \cite{mosel}, yielding 
the  following form:
\begin{eqnarray} \label{S0}
 {S}_0
= \tilde{S}_{0}
 ( i \gamma \cdot \partial + M) ,
\end{eqnarray}
where
$\tilde{S}_{0}  = \tilde{S}_{0}  (k) \equiv \frac{1}{k^2 - {M}^2}$.
 To write down all the chirally symmetric terms up to the fourth
order the properties of the traces  in flavor  and  Dirac matrices 
(\ref{properties-4}) were used.
These  terms up to the fourth order in the expansion
of the determinant, always in the zero order derivative expansion,
and up to the second order in the quark mass,
for the chiral  symmetry breaking 
 but  isospin invariant terms,
are given by:
\begin{eqnarray} \label{CHPT}
{\cal L}_{\pi}
 &=&
\frac{a_1}{2} tr  \;
\partial_\mu U 
\partial^\mu   U^\dagger
+ 
a_{sb} tr \;  M  (U + U^\dagger )
\nonumber
\\
&&
+
l_1 
\;  (tr \partial_\mu U \partial^\mu U^\dagger )^2
+ 
l_2  
\; tr (\partial_\mu U^\dagger \partial_\nu U) 
tr (\partial^\mu U^\dagger \partial^\nu U)
\nonumber
\\
&& 
+ 
l_3 
\; 
 tr \;  [ M  ({U^\dagger} + U)  ]
\;
 tr \; [\partial^\mu U \partial_\mu U^\dagger ]
\nonumber
\\
&& +  l_4
\;
 ( tr \;  [ M  (U^\dagger + U ) ]  )^2 ,
\end{eqnarray}
where the trace $tr$ stands for trace in  flavor indices.
By resolving Dirac and color indices traces,
the low energy coefficients were defined as:
\begin{eqnarray}
a_1 g_{\mu\nu} &=& 
2  i^3   \; N_c  \; 
Tr''  \tilde{S}_0^2 F^2 \gamma_\mu \gamma_\nu
,
\nonumber
\\
a_{sb} 
 &=&
4 i N_c  \; Tr'' \; \tilde{S}_0  F
,
\nonumber
\\
l_1  \Gamma_{\mu\nu\rho\sigma}
&=&  - 2 \;
  l_2    \Gamma_{\mu\nu\rho\sigma} \;  =  \;
\frac{ i^5}{2}
 N_c \; Tr'' \; 
\tilde{S}_0^4 F^4
 \gamma_\mu \gamma_\nu
 \gamma_\rho \gamma_\sigma
,
\nonumber
\\
l_3  g_{\mu\nu} 
&=& 
i \; N_c \; Tr'' \; \tilde{S}_0^3 F^3
 \gamma_\mu \gamma_\nu
,
\nonumber
\\
l_4   &=& 
-\frac{ i}{2}  N_c  \;
 Tr'' \; \tilde{S}_0^2  F^2
,
\end{eqnarray}
where the traces in flavor were not included in $Tr''$.
However it is important to note that, 
as shown in Refs. \cite{PRC,E-R1986},
by considering the leading terms of the derivative expansion 
with the full meson  form factors, a different structure appears.
In particular, for  the  following  second order 
term of the determinant  expansion 
$\partial_\mu U (x) \partial^\mu U^\dagger (y)$,
that  can be expanded  in a  derivative expansion.
Consider a non trivial form factor ($G(p^2)$ , non punctual pion) with which it 
can be written as:
\begin{eqnarray} \label{der-expansion}
&& Tr \partial_\mu U(y) \partial^\mu U^\dagger (x) 
G(p^2)
\nonumber
\\
&\sim&
Tr \left\{  \partial_\mu U(x) 
+ 
(y-x)^\nu \partial_\nu \partial_\mu U (x) 
+
\frac{1}{2} (y-x)^\rho (y-x)^\nu  \partial_\rho \partial_\nu \partial_\mu U (x)
\right\}   \partial^\mu U^\dagger (x) G(p^2)
 ,
\end{eqnarray}
where
the first order term of this expansion is zero 
because 
$
Tr \; (y-x)_\nu G(p^2)  \sim i\;  Tr \; \left. \frac{\partial}{\partial p}
 G(p^2) \right|_{p=0} =0.
$
The second order term in expression (\ref{der-expansion}) after an integration by parts
can be written as:
\begin{eqnarray} \label{last-4th}
I_2 = - \; tr_F \; G_{2,U} \; \partial^2 U \;\partial^2 U^\dagger
\end{eqnarray}
where, by resolving the trace as integration in momenta,:
$
G_{2,U} = i 2 N_c \int_{p,q}\frac{\partial^2 G(p^2)}{\partial p^2}.
$
This term completes the leading terms of Chiral Perturbation Theory.

\subsection{ Complete second order effective model }

Expressions
(\ref{deriv-1},\ref{freePI},\ref{2nd-I},\ref{sigma-term},\ref{I-4},\ref{2nd-sb})  and the constituent
quark and gluon free   terms in expression  (\ref{free-q}), with the 
calculated corrections to the pion mass and quark effective mass  (\ref{Meff}), 
are written below together.
With the field rescaling discussed above, 
  the second order terms of the expansion it yields:
\begin{eqnarray} \label{Lcomplete}
{\cal L}_{eff}^{(2)} &=&
t_1  \bar{\psi}_1 \left(  i \gamma \cdot \partial
- M^* \right) \psi_1
- 2 t_1  g_A  ({\cal D}_\mu {\pi}_i) 
   \bpsi   i \gamma_5 \gamma^\mu \sigma_i \psi
\nonumber
\\
&+&  t_1  g_{\pi d \pi}
\frac{\vec{\pi} \times \partial^\nu \vec{\pi}}{1+{\vec{\pi}}^2}  
\cdot (\bpsi   \gamma_\nu  \vec{\sigma} \psi)
 + \frac{k_2}{2}
 {\cal D}_\mu \pi_i  {\cal D}^\mu \pi_i 
\nonumber
\\
&-&
 c_1 ({\cal D}_\mu \pi_i \cdot {\cal D}^\mu \pi_i)^2 
- c_2  ({\cal D}_\mu \pi_i \cdot  {\cal D}_\nu \pi_i)^2 
\\
&+& 
 {\cal L}_{sb} 
 +  I_{h.o.d.} 
- 
t_1^2 
\frac{g^2 }{2} \int_y j_{\mu}^b (x)  \tilde{R}^{\mu \nu}_{bc} (x,y) j_{\nu}^{c} (y) 
\nonumber
\\
&+&
t_1^2 
\left\{   g_{NJL}^{h.o.}  \left[ ( \bpsi  \psi )^2
 +   (  \bpsi \sigma_i \gamma_5 \psi )^2  \right]
+
g_{vNJL}
\left[ ( \bpsi  \sigma_i \gamma_\mu \psi )^2
 + ( \bpsi \sigma_i \gamma_\mu \gamma_5 \psi )^2
\right]
\right\} ,
\nonumber
\end{eqnarray}
where ${\cal L}_{sb}$ are the symmetry breaking terms up to the second order in the 
current quark mass, including (\ref{2nd-sb}),:
\begin{eqnarray} \label{ISB}
{\cal L}_{sb} &=&
c_{\sigma,sb} t_1
 \frac{\vec{\pi}^2}{1 + \vec{\pi}^2} ( \bpsi \psi )
-
 M_\pi^2 F^2
\frac{\vec\pi^2}{1+ \vec{\pi}^2}
+ V_2
\frac{\vec\pi^4}{(1+ \vec{\pi}^2)^2}
\nonumber
\\
&+&
\frac{ V_{3B}  \vec\pi^2}{(1+ \vec{\pi}^2)}
\left( {\cal D}^\mu \vec{\pi} \cdot {\cal D}_\mu \vec{\pi} \right)
-
 V_{3C}
\frac{
\left( \vec{\pi} \cdot {\cal D}^\mu \vec{\pi}  \right)^2
}{(1+ \vec{\pi}^2)}
\nonumber
\\
&+& 
\frac{V_{4B}  (\vec\pi^2)^2}{(1+ \vec{\pi}^2)^2}
\left( {\cal D}^\mu \vec{\pi} \cdot {\cal D}_\mu \vec{\pi} \right)
+V_{4C} 
\frac{  (\vec{\pi} \cdot {\cal D}^\mu \vec{\pi})
(\vec{\pi} \cdot {\cal D}_\mu \vec{\pi})
}{(1+ \vec{\pi}^2)^2} 
,
\end{eqnarray}
$I_{h.o.d.}$ stands for higher oder terms
in momentum and in the derivative expansion.
The (constituent) quark mass was defined as $M^* = m + \Delta  m^*$,
which is therefore not necessarily the same contribution as
 the effective mass $m^*$ for the quarks that constitutes 
pions  and the scalar condensate 
which can be 
determined from  a gap equation.
The 
terms with the 
effective parameters calculated above, such as  
 $\Delta m^*$, $g_A$, $g_{\pi d \pi},M_\pi^2,c_{\sigma,sb}$,  $k_2$
and $V_{i}$
 depend on the parameter $t_2$, that was set equal to $1$, and  which 
indicates the quark component that was  integrated out.
The first  three  lines of expression
 (\ref{Lcomplete})  correspond to the  
large $N_c$ constituent quark and pion 
effective theory  proposed in Ref. \cite{Weinberg2010}
which however can receive corrections from the derivative expansion 
of the quark determinant.
The  effective quark-quark
interactions and the symmetry breaking pion-quark coupling 
($c_{\sigma,sb}$) and higher order pion interactions
are  of higher order in the $1/m^*$ (and also  $1/N_c$) expansion.

\section{Summary}

In this work  an 
effective model  for quarks and pions  was derived  
from the global color model (\ref{Seff}).
In the chiral limit, it corresponds to the large $N_c$ effective theory proposed by Weinberg
\cite{Weinberg2010}.
 For that,
the quark field was splitted  into two components by means of the one loop
background field method.
 The component
that yields light quark-antiquark  mesons and the chiral condensate
was integrated out
 by means of the auxiliary field method.
The  one-loop level is obtained 
 by performing the shift of all the quark bilinears
which also 
is compatible with considering  quark-antiquark   light meson states.
Therefore no criterium  of the type of 
low and high energy modes was considered.
However,
it was noticed that this might involve  an ambiguity in performing such 
splitting the quark field and this
gives rise to a possible
 ambiguity in  eventual   contributions  of the 
Skyrme terms  and constituent (background) 
quarks to the 
 baryon  structure.
Since the structure of the results are independent of the separation criterium,
 this is left for further investigations.
This can be seen in expression (\ref{Lcomplete})
where the constituent quark contribution has the parameter $t_1$ whereas
the Skyrme terms have implicitly  $t_2^2=1$.
The vector auxiliary fields that give rise to the local lightest
vector meson fields were neglected  because these 
excitations are 
considerably heavier than the pion 
and only contribute in relatively higher energies 
than the low energy regime 
dictated by pions.
However, constituent quarks and  pions
are  also  expected to contribute for vector meson
structure and dynamics  and
 the analysis of the corresponding contribution
  is outside the scope of the work.
Nevertheless it  is interesting to remember
that the fourth order (Skyrme) terms above,  or corrections to them,
might be obtained 
from the vector meson dynamics \cite{vector-meson-skyrme}.
The part of this model that involves
vector mesons interactions with quarks  will be presented elsewhere.
A chiral rotation has been performed 
for  the limit of local structureless light mesons
 (pions) which yielded a particular 
dependence of the results on covariant derivatives 
of the quark and pion fields.
Should the calculations be carried out  
without the structureless limit,
some of the resulting expressions  for the effective couplings or 
mass parameters
would yield Schwinger-Dyson or  Bethe-Salpeter like equations.
Nevertheless  results 
reproduce correctly the
expected terms from Chiral Perturbation Theory for punctual pions,
the expected leading  pion-quark effective interactions 
as well as the GellMann Oakes Renner relation as the leading symmetry breaking
low energy relation.
A longwavelength and weak quark field 
expansion was
performed by considering the local punctual meson fields 
in  the zero order derivative expansion.
The determinant expansion is also equivalent to a large quark 
effective mass $m^*$ expansion or  small coupling constant $g^2$.
For that, two different definitions of the pion field were considered
which are related by chiral rotations from the original linear realization
 of chiral symmetry.
In one of these pion definitions, the chiral invariant pion terms are always
written in terms of the covariant pion derivative (${\cal D}_\mu \vec{\pi}$)
which allowed to compare the final expressions to
the effective  theory proposed in Ref. \cite{Weinberg2010}.
With the redefinition of the pion and quark fields a non trivial
 Jacobian appears in the functional measure 
of the generating functional which  was not exhibited and discussed in this work
because it involves higher order quark-pion interactions, of the 
form ${\cal O}(\bpsi \zeta \psi (\vec{\pi})^n)$.
The pion sector was rewritten in terms of a second pion definition
showing that it turns out 
to provide basically the leading terms of  chiral perturbation theory.
The corresponding 
expressions for the low energy constants were exhibited.
The effective  coupling constants
 and parameters of the effective model (\ref{Lcomplete})
were  expressed in terms of the 
 coupling $g^2$ and  quark current 
and constituent, $m^*$, masses.
All the derivation was carried out without specifying the gluon propagator what 
 would be needed to provide  numerical estimations. 
The present approach 
  makes  possible
to introduce further systematic corrections
 in the  expansions that were performed.
By closing external legs 
the higher order terms of the expansion
 yield loop corrections for the effective  couplings and parameters
 above 
with progressively large powers of $1/m^*$.
Corrections   for intermediary, or 
relatively larger, energies  might also be considered 
  in the derivative expansion and  by  considering the full 
meson form factors.
 Finally, higher order corrections to the background field method can also be envisaged.

\section*{Acknowledgments}

The author thanks  short discussions with 
G.I. Krein
and C.D. Roberts,
and partial financial support by CNPq- Brazil.


\end{document}